\documentstyle [11pt,epsfig]{article}
\begin{document}

{\Large{\bf{
Measurement of {\boldmath $xF_3$} and {\boldmath $F_2$} Structure
 Functions in Low {\boldmath $Q^2$} Region
 with the IHEP-JINR Neutrino Detector}
}}

\vspace{0.5cm}
A.~V.~Sidorov$^a$, V.~B.~Anykeyev$^b$,
Y.~A.~Batusov$^a$, S.~A.~Bunyatov$^a$, A.~A.~Borisov$^b$,
N.~I.~Bozhko$^b$, S.~K.~Chernichenko$^b$, G.~L.~Chukin$^b$,
O.~Y.~Denisov$^b$, R.~M.~Fachrutdinov$^b$, V.~N.~Goryachev$^b$,
M.~Y.~Kazarinov$^a$, M.~M.~Kirsanov$^b$, O.~L.~Klimov$^a$,
A.~I.~Kononov$^b$, A.~S.~Kozhin$^b$, A.~V.~Krasnoperov$^a$,
V.~I.~Kravtsov$^b$, V.~E.~Kuznetsov$^a$, V.~V.~Lipajev$^b$,
A.~I.~Mukhin$^b$, Y.~A.~Nefedov$^a$, B.~A.~Popov$^a$,
S.~N.~Prakhov$^a$, Y.~I.~Salomatin$^b$, V.~I.~Snyatkov$^a$,
Y.~M.~Sviridov$^b$, V.~V.~Tereshchenko$^a$, V.~L.~Tumakov$^b$,
V.~Y.~Valuev$^a$, A.~S.~Vovenko$^b$ \\ [10mm]

 ~$^a$ JINR, 141980 Dubna Moscow Region Russia\\

\vspace{0.3cm}
 ~$^b$  IHEP, 142284 Protvino Moscow Region Russia\\
\vspace{1.5cm}

\begin{abstract}
The isoscalar structure functions $xF_3$ and $F_2$ are measured
as functions of $x$ averaged over all $Q^2$ permissible for the
range of 6 to 28~GeV of incident neutrino (anti-neutrino) energy
at the IHEP-JINR Neutrino Detector.
The QCD analysis of $xF_3$ structure function provides
$\Lambda_{\overline{MS}}^{(4)} = (411 \pm 200)$~MeV
under the assumption of QCD validity in the region of low $Q^2$.
The corresponding value of the strong interaction constant
$\alpha_S (M_Z) = 0.123^{+0.010}_{-0.013}$
agrees with the recent result of the CCFR collaboration
and with the combined LEP/SLC result.
\end{abstract}

%
\newpage
\par
\vspace{1cm}

\section{Introduction}\label{sec:intro}

The data on deep-inelastic neutrino and anti-neutrino scattering
in a wide region of momentum transfer provide a reliable basis
for precise verification of QCD predictions~\cite{Altarelli}.
In this paper we present the measurements of the $xF_3$ and $F_2$
structure functions (SF) and the QCD analysis of $xF_3$
in the kinematic region of relatively small
momentum transfer $0.55<Q^2<4.0$~GeV$^2$. The
value
of the strong
interaction constant $\alpha_S\,(M_Z)$
is
also evaluated and compared with the results
of other experiments.

\section{Data Samples}\label{sec:data}

The analysis is based on data collected with three independent exposures
of the IHEP-JINR Neutrino Detector~\cite{ND} to the wide-band neutrino
and anti-neutrino beams~\cite{beams} of the Serpukhov U-70 accelerator.
The exposure to the anti-neutrino beam ($\overline\nu_\mu$-exposure)
was performed at the proton beam energy $E_p=70$~GeV, whereas the two
$\nu_\mu$-exposures were carried out at $E_p=70$~GeV and at $E_p=67$~GeV.
The energy of the resulting $\nu_\mu$ ($\overline\nu_\mu$) was in the range
of $6 < E_{\nu(\overline\nu)} < 28$~GeV.

The experimental set-up and the selection criteria of charged current (CC)
neutrino and anti-neutrino interactions are discussed in~\cite{publ}.
We restricted the range of measurements to $W^2 > 1.7$~GeV$^2$ in order
to reject quasi-elastic and resonance events and
select mainly deep-inelas\-tic neutrino and anti-neutrino interactions.
The number of protons on target (p.o.t.) for each exposure, the selected
number of $\nu_\mu$~CC and $\overline\nu_\mu$~CC events and the mean values
of $Q^2$ for the three data samples are given in Table~\ref{tab:stat}.
\begin{table}[ht]
\begin{center}
\caption{Summary of the exposures.}\label{tab:stat}
{\normalsize{
\begin{tabular}{l|ccc} \hline
Beam                             & $\overline\nu_\mu$ &
                                   $\nu_\mu$ & $\nu_\mu$ \\
$E_p$, GeV                       & ~~$70$~ & ~$70$~ & ~$67$~\\
$N_{p.o.t.}$ $\times$ 10$^{17}$  &   2.86  &  1.05  &  2.11 \\
Final statistics                 & ~~$741$~ & ~$2~139$~ & ~$3~848$~\\
$\langle Q^2 \rangle$, GeV$^2$~~ & ~$1.2$ &
                 \multicolumn{2}{c}{~$2.3$}\\ \hline
\end{tabular}
}}
\end{center}
\end{table}

\section{Data Analysis}\label{sec:anal}

    The SF were measured as functions of $x$ averaged over all $Q^2$
permissible for the energy range $6 < E_{\,\nu\,(\overline\nu)} < 28$~GeV.
The events were binned in intervals of $x$, and the values of $xF_3$
and $F_2$ were calculated in these intervals.

    The number of $\nu_\mu$ interactions, $n^\nu$, and $\overline\nu_\mu$
interactions, $n^{\overline\nu}$, in a given bin of $x$ is a linear
combination of the average values $\{F_2\}$ and $\{xF_3\}$ of the
respective SF in this bin (we assume invariance under the charge
conjugation):
\begin{eqnarray*}
  n^{\,\overline\nu} & = & a^{\,\overline\nu}\cdot \{F_2\} ~-~
   b^{\,\overline\nu}\cdot\{xF_3\},\\
  n^{\,\nu}_{\,1,2}  & = & a^{\,\nu}_{\,1,2}\cdot \{F_2\} ~+~
   b^{\,\nu}_{\,1,2} \cdot \{xF_3\}.
\end{eqnarray*}
The subscripts $1$ and $2$ correspond to the $\nu_\mu$-exposures at
$E_p=70$~GeV and $E_p=67$~GeV respectively. The quantities
$a^{\,\nu\,(\overline\nu)}$ and $b^{\,\nu\,(\overline\nu)}$ are the
integrals (``flux integrals'') of products of the differential neutrino
(anti-neutrino) flux $\phi^{\,\nu\,(\overline\nu)}\,(E)$ and the known
factors depending on the scaling variables $x$ and $y$, as given by
the standard form of the differential cross-section for deep-inelastic
$\nu_\mu\,(\overline\nu_\mu)$-scattering off an isoscalar target:
\begin{eqnarray*}
  a^{\,\nu\,(\overline\nu)} & = & N\, \frac{G^2M}{\pi}\times \\
    & & \int_{}^{}(1-y-\frac{Mxy}{2E}+\frac{y^2}{2\,(R+1)})\,
    E\,\phi^{\,\nu\,(\overline\nu)}(E)\,dx\,dy\,dE\,, \\
  b^{\,\nu\,(\overline\nu)} & = &
    N\, \frac{G^2M}{\pi}
    \int_{}^{}y\,(1-\frac{y}{2})\,E\,\phi^{\,\nu\,(\overline\nu)}
    (E)\,dx\,dy\,dE\,.
\end{eqnarray*}
Here $N$ is the number of nucleons in the fiducial volume
of the detector and the parameter $R=(F_2-2xF_1)\Big/2xF_1$ measures
the violation of the Callan-Gross relation~\cite{callan}.

The number $n^{\,\nu(\overline\nu)}$ of neutrino (anti-neutrino)
interactions in a given bin of $x$ was obtained from the measured number
of neutrino (anti-neutrino) events in this bin corrected for acceptance,
smearing effects arising from Fermi motion and measurement
uncertainties, radiative effects (following the prescription given
by De R\'ujula et al.~\cite{rujula}) and target non-isoscalarity
(assuming $d_v/u_v=0.5$)~\cite{noni}. To determine appropriate
correction factors, the Monte Carlo simulation of the experimental
set-up has been carried out using the CATAS program~\cite{catas}.
We used the Buras and Gaemers (BEBC) parameterization~\cite{buras} for
quark distributions. The charm quark content of the nucleon was
assumed to be zero. The kinematic suppression of $d \rightarrow c$
and $s \rightarrow c$ transitions was taken into account assuming
slow rescaling~\cite{slow} and the charm and strange quark masses
of $m_c=1.25$~GeV and $m_s=0.25$~GeV respectively. The Fermi motion of
nucleons was simulated according to~\cite{fermi}. The details of the
Monte Carlo simulation are described in~\cite{publ,blum}.

The number of interactions in a given bin of $x$ is subjected to
kinematic constraints imposed by the cuts on the muon momentum
($p_\mu$~$>$~1~GeV/$c$~\cite{publ}), on the neutrino (anti-neutrino)
energy ($6<E_{\,\nu (\overline\nu)}<28$~GeV) and on the invariant mass
square of the hadronic system ($W^2 > 1.7$~GeV$^2$). These constraints
were taken into account in the calculation of the flux integrals by
appropriate modification of the volume of integration.

The measured values of $F_2$ and $xF_3$ structure functions are
given in Table~\ref{tab:sfun} and in Fig.~1.
\begin{table*}[thb]
\begin{center}
\caption{The isoscalar structure functions $F_2$ and $xF_3$ obtained
 with the assumption of $R = 0$. The difference $\Delta F_2$ between the
 values of $F_2$ obtained with $R = 0.1$ and with $R = 0$ is also presented.
 The bin edges are at $x$ = 0.02, 0.1, 0.2, 0.3, 0.4, 0.5 and 0.65. The
 shown systematic errors do not include the normalization error of 4\%
 for $F_2$ and 11\% for $xF_3$ originating from the uncertainties in the
 $\nu_\mu$ and $\overline\nu_\mu$ flux prediction~\cite{flux}.}
\label{tab:sfun}
\vspace*{0.5cm}
{\normalsize{
\begin{tabular}{c|c|rccc|ccc} \hline
$\langle x \rangle$ & $\langle Q^2 \rangle$, GeV$^2$ & $F_2$~~ & $stat$ &
         $syst$ & $\Delta F_2$~ & $xF_3$ & $stat$ & $syst$ \\ \hline
~$0.052$~ & $0.55$ & ~$1.169$ & ~$.039$~ & ~$0.047$~ & ~$0.023$~~ &
         ~$0.445$~ & ~$.458$~ & ~$0.062$~ \\
~$0.148$~ & $1.4$  &  $1.097$ &  $.036$  &  $0.022$  &  $0.022$  &
          $0.583$  &  $.087$  &  $0.017$ \\
~$0.248$~ & $2.2$  &  $0.894$ &  $.032$  &  $0.018$  &  $0.019$  &
          $0.622$  &  $.075$  &  $0.019$ \\
~$0.346$~ & $2.9$  &  $0.576$ &  $.028$  &  $0.017$  &  $0.013$  &
          $0.556$  &  $.109$  &  $0.011$ \\
~$0.447$~ & $3.4$  &  $0.390$ &  $.025$  &  $0.012$  &  $0.009$  &
          $0.336$  &  $.070$  &  $0.007$ \\
~$0.563$~ & $4.0$  &  $0.182$ &  $.017$  &  $0.004$  &  $0.004$  &
          $0.177$  &  $.117$  &  $0.005$ \\ \hline
\end{tabular}
}}
\end{center}
\end{table*}

The systematic errors presented in Table~\ref{tab:sfun} come from the
uncertainties in the knowledge of neutrino flux and
cross-sections~\cite{publ}, imperfect detector calibration and uncertainties
in the correction factors due to the choice of the input quark
distributions. The overall normalization error, originating from
the uncertainties in the $\nu_\mu$ and $\overline\nu_\mu$ flux
prediction~\cite{flux}, was estimated to be 4\% for $F_2$ and
11\% for $xF_3$. The correction factor uncertainties were evaluated
by repeating the calculation of the SF using the
Field-Feynman~\cite{field} and GRV~\cite{grv} parameterizations of
quark distributions.

\begin{figure}[th]
\vspace{1mm}
\begin{center}

\epsfxsize=10cm
\epsfysize=10cm
  \vspace*{-.5cm}

{\bf
\parbox[h]{15.cm}{\baselineskip  4pt
\centerline{\epsfbox{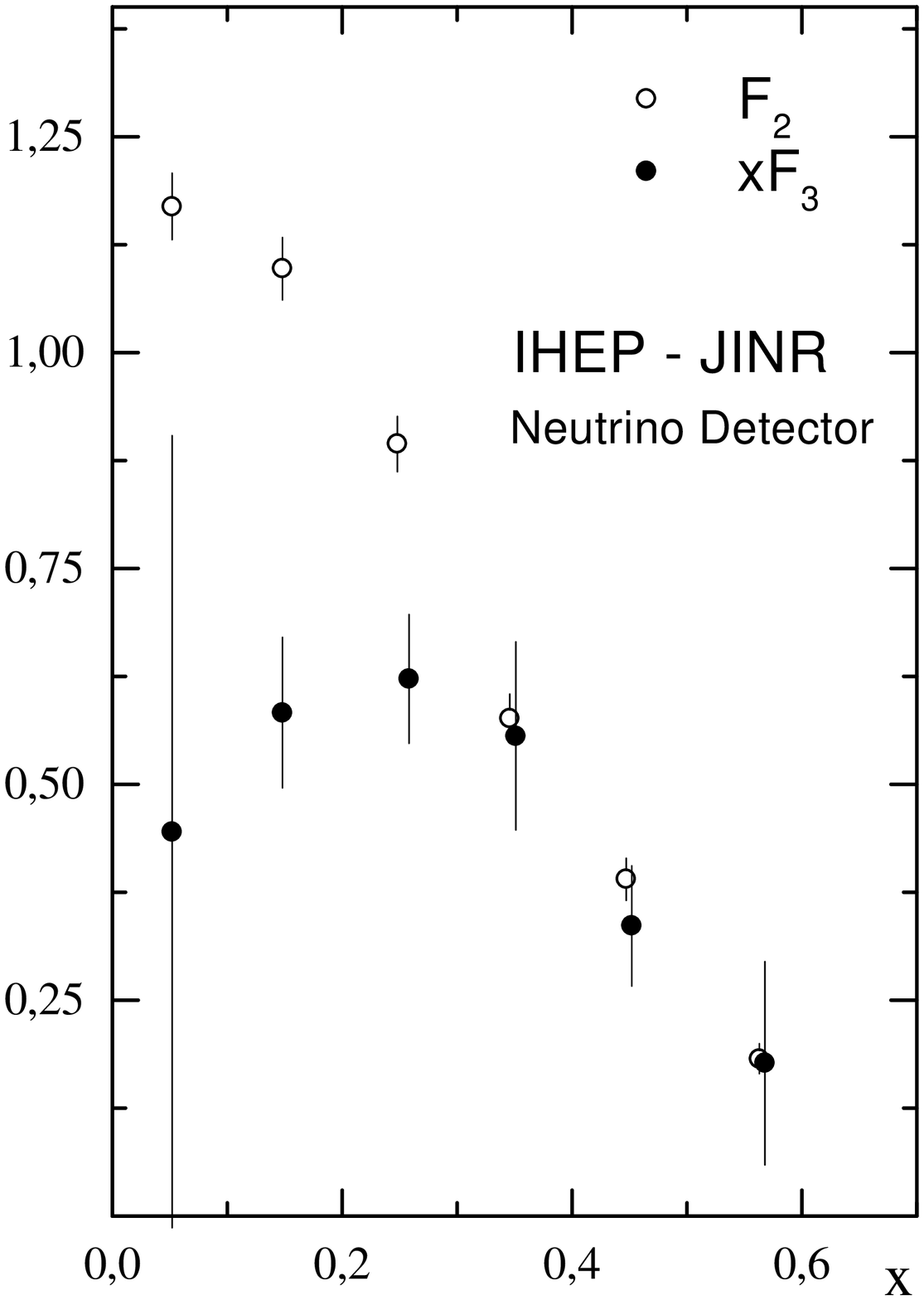}}
\caption{
 The measured $x$-dependence of the isoscalar structure functions
  $F_2(x)$ and $xF_3(x)$. The statistical and systematic errors are
  added in quadrature (the normalization errors of $4\%$ for $F_2$
  and $11\%$ for $xF_3$ are not shown).
}
}}\\[15mm]
\end{center}
\end{figure}

The obtained experimental data on the $xF_3$ were then compared with the QCD
prediction for $Q^2$-evolution by the Jacobi polynomials method in the
next-to-leading order (NLO) QCD approximation \cite{JacobiKriv,Kriv2,KaSi}.
Performing the QCD analysis of the $xF_3$ SF, we do not discuss here
the problem of validity of application of perturbative QCD predictions
for the kinematic region of low $Q^2$ and do not take into account nuclear
effects, heavy quarks threshold effects and higher order QCD corrections.

In order to take into account the target mass corrections, the
Nachtmann moments~\cite{Nacht} of $F_3$ are expanded in powers
of $M_{nucl}^2/Q^2$. Retaining only the terms of the order of
$M_{nucl}^2/Q^2$, one could obtain:
\begin{eqnarray}
  M_{3}(N,Q^2) & = & M_{3}^{QCD}(N,Q^2)\,+\, \label{m3}\\
    & & \frac{N(N{+}1)}{N+2}\,\frac{M_{nucl}^2}{Q^2}
        \,M_{3}^{QCD}(N+2,Q^2).\nonumber
\end{eqnarray}
Here $M_3^{QCD}(N,Q^2)$ are the Mellin moments of $xF_3$ :

\begin{eqnarray}
M_3^{QCD}(N,Q^2)&=&\int_{0}^{1}{x^{N-2}}xF_{3}(x,Q^2)\,dx, \label{Mellf}\\
 ~~~N = 2,3, \dots
\nonumber
\end{eqnarray}

The $Q^2$-evolution of Mellin moments is defined by QCD \cite{s4,s5}
and is presented here for the non-singlet case for simplicity:
\begin{eqnarray}
  M_3^{QCD}(N,Q^2) & = & \left[\frac{\alpha _{S}\left(Q_{0}^{2}\right)}
    {\alpha _{S}\left(Q^{2}\right)}\right]^{d_{N}}H_{N}
    \left(Q_{0}^{2},Q^{2}\right)\times \nonumber\\
  & & M_3^{QCD}(N,Q^2_0),~~~N = 2,3, \dots \label{m3q2} \\
  d_N & = & \gamma^{(0)NS}_{N}\bigg/2\beta_0, \nonumber\\
  \beta_0 & = & 11-\frac{2}{3}n_f. \nonumber
\end{eqnarray}
Here $\alpha_s(Q^2)$ is the strong interaction constant, $\gamma^{(0)NS}_{N}$
are the non-singlet leading order anomalous dimensions and $n_f$ is the number
of flavours. The factor $H_{N}\left (Q_{0}^{2},Q^{2}\right)$ contains all
next-to-leading order QCD corrections~\cite{KaSi,s5,KKPS}.
\newpage
The unknown coefficients $M_3^{QCD}(N,Q^2_0)$ in (\ref{m3q2}) could be
parameterized as the Mellin moments of some function:
\begin{eqnarray}
  M_3^{QCD}(N,Q^2_0) & = &\int_{0}^{1}{x^{N-2}}Ax^b(1-x)^c\,dx,
  \label{Mellf30}\\
  & & ~~~N = 2,3,\dots \nonumber
\end{eqnarray}
where the constants $A$, $b$ and $c$ should be determined from the fit to
the data. Having defined the moments (\ref{m3})$\,$--$\,$(\ref{Mellf30})
and following the method discussed in~\cite{JacobiKriv,Kriv2},
we can write the $xF_3$ SF in the form:
\begin{eqnarray*}
  xF_{3}^{QCD}(x,Q^2) & = & x^{\alpha}(1-x)^{\beta}
        \sum_{n=0}^{N_{max}}\Theta_n^{\alpha,\beta}(x)\times\\
  & & \sum_{j=0}^{n}c_{j}^{n}{(\alpha,\beta )}
         M_3\left(j+2,\,Q^{2}\right),
\end{eqnarray*}
where $\Theta^{\alpha,\beta}_{n}(x)$ are the Jacobi polynomials
and $c^{n}_{j}(\alpha,\beta)$ are the coefficients of the
expansion of $\Theta^{\alpha,\beta}_{n}(x)$ in powers of $x$:
$$
\Theta_{n} ^{\alpha , \beta}(x)=
\sum_{j=0}^{n}c_{j}^{n}{(\alpha ,\beta )}\,x^j\,.
$$

The accuracy of the SF approximation better than 1\% is achieved
for $N_{max} = 9$ in a wide region of the parameters $\alpha$ and
$\beta$~\cite{Kriv2}.

The higher-twist (HT) contribution is also taken into account:
$$
xF_{3}(x,Q^2)=xF_{3}^{QCD}(x,Q^2)+\frac{h(x)}{Q^2},
$$
where $h(x)=0.166-3.746\cdot x+9.922\cdot x^2-6.730\cdot x^3$ is chosen
by an interpolation of the NLO result for the HT contribution
from~\cite{Sasha}. This shape of $h(x)$ is in a good agreement
with the theoretical prediction of~\cite{webber} and with the result
of~\cite{Siht,KKPS2} obtained for a higher $Q^2$ kinematic region.
\begin{table}[ht]
\begin{center}
\caption{The results of the NLO QCD fit to the $xF_3$ SF data for $n_f=4$,
 $Q^2_0\,=\,3$~GeV$^2$, $N_{max}=9$, $\alpha=0.7$, $\beta=3.0$.}
\vspace*{0.5cm}
{\normalsize{
\begin{tabular}{cc}        \hline
     $\chi^2$                   &   0.22                  \\ \hline
       $A$                      &   10.4~~(fixed)         \\
       $b$                      &   0.86 $\pm$ 0.14       \\
       $c$                      &   3.83 $\pm$ 0.61       \\
$\Lambda_{\overline{MS}}^{(4)}$ & ~~(411 $\pm$ 200) MeV~~ \\ \hline
~~$\alpha_S (M_Z)$~~            & $0.123^{+0.010}_{-0.013}$ \\ \hline
\end{tabular}
}}
\label{tab:res}
\end{center}
\end{table}

Using nine Mellin moments
and taking into account target mass corrections, we have determined
four parameters -- $A$, $b$, $c$ and the QCD parameter
$\Lambda_{\overline{MS}}$ (Table~\ref{tab:res}).
In order to decrease the number of free parameters we have fixed
the value of parameter $A$ using the Gross-Llewellyn Smith sum rule
$Q^2$ - dependence: $S_{GLS}=3~(1-\frac{\alpha_{s}(Q^2_0)}{\pi})$~\cite{gls}.
The fit was
performed using the MINUIT program~\cite{minuit}. Three sources
of errors -- statistical, systematic and normalization -- were
summed up in quadrature. The errors for the free parameters corresponding
to the $70\%$ confidence level were obtained using the procedure
described in~\cite{minuit_errors}. A relatively good accuracy
of the measurements of $\Lambda_{\overline{MS}}$ was achieved
due to a high sensitivity of the QCD evolution equations to the
variations of $\Lambda_{\overline{MS}}$ in the low $Q^2$ region
($0.55<Q^2<4.0$~GeV$^2$ in our case).

The value of $\alpha_S (M_Z)$ corresponding to the measured value
of $\Lambda_{\overline{MS}}$ was calculated from the so-called
``matching relation''~\cite{marc} and found to be
$\alpha_S (M_Z) = 0.123^{+0.010}_{-0.013}$.

\section{Discussion of the results}\label{sec:disc}

We have compared our results with the measurements performed by other
experiments. The comparison led to the following comments:
\begin{itemize}
\item The parameter
$\Lambda_{\overline{MS}}^{(4)}$ = ~~(411 $\pm$ 200) MeV~~
      of the fit to the $xF_3$ SF data
      is in agreement
      with the NLO analyses with HT contribution of the CCFR $xF_3$ data:
      $\Lambda_{\overline{MS}}^{(4)} = (381\pm 53\,(stat)\pm\,
      17(HT))$~MeV~\cite{CCFRrev} and
      $\Lambda_{\overline{MS}}^{(4)} = (428\pm 158\,(exp))$~MeV~\cite{KKPS2}.
\item The value of $\Lambda_{\overline{MS}}^{(4)}$ obtained from the NLO
      analysis of the $xF_3$ SF provides the value of the strong interaction
      constant at the point of Z boson mass of
$\alpha_S (M_Z) = 0.123^{+0.010}_{-0.013}$
      which is in agreement with the result of the analysis of the
      CCFR data $\alpha_s(M_Z) = 0.119 \pm 0.002\,(exp) \pm 0.004\,
      (theory)$~\cite{CCFRrev} and with the combined LEP/SLC result
      $\alpha_s(M_Z) = 0.124 \pm 0.0043$~\cite{RPP}.
      Our current measurement is
higher
by one standard deviation
than the value
      $\alpha_s(M_Z) = 0.113 \pm 0.003\,(exp) \pm 0.004\,(theory)$~\cite{VM},
      obtained in the $F_2$ structure function analysis of the BCDMS and
      SLAC data on $\mu N$ and $e N$ deep-inelastic scattering.
\end{itemize}

\section{Conclusion}\label{sec:concl}

We have presented the measurements of the structure functions $F_2$ and
$xF_3$ in the kinematic range $0.02 < x < 0.65$ and $0.55 < Q^2 < 4.0$~GeV$^2$,
obtained from the inclusive deep-inelastic $\nu_\mu$ and $\overline{\nu}_\mu$
scattering data collected at the IHEP-JINR Neutrino Detector.
The NLO QCD analysis with HT contributions of $xF_3$
under the assumption of QCD validity in the region of low $Q^2$
provides $\Lambda_{\overline{MS}}^{(4)}$ = ~~(411 $\pm$ 200)~MeV;
the corresponding value of the strong interaction constant is
$\alpha_S (M_Z) = 0.123^{+0.010}_{-0.013}$.

\section*{Acknowledgments}

This work has been supported by the Russian Foundation for Basic Research
under grants 96-02-17608, 96-02-18562 and 99-01-00091.

\end{document}